\documentclass[conf, 10pt]{IEEEtran}

\usepackage{cite}
\usepackage{amsmath,amssymb,amsfonts,amsthm, empheq}
\usepackage{algorithm}
\usepackage{algorithmic}
\usepackage{setspace}
\usepackage{graphicx}
\usepackage{textcomp}
\usepackage{xcolor}
\usepackage{mathtools}
\usepackage{graphicx}
\usepackage{booktabs}
\usepackage{makecell}
\usepackage[font={small}]{caption}
\usepackage{subcaption}
\usepackage{cite}
\usepackage{breqn}
\usepackage{mathrsfs}
\usepackage{accents}
\usepackage{acronym}
\usepackage{multirow}
\usepackage{soul}
\usepackage{bm}
\usepackage{url}
\usepackage{wrapfig}
\usepackage{todonotes}
\usepackage{verbatim}
\usepackage{pdfpages}
\usepackage{siunitx}
\usepackage{adjustbox}
\usepackage{steinmetz}
\usepackage{enumitem}

\usepackage{glossaries}

\newacronym{3gpp}{3GPP}{Third Generation Partnership Project}
\newacronym{fr3}{FR3}{Frequency Range 3}
\newacronym{fr2}{FR2}{Frequency Range 2}
\newacronym{fr1}{FR1}{Frequency Range 1}
\newacronym{6g}{6G}{Sixth Generation}
\newacronym{5g}{5G}{Fifth Generation}
\newacronym{4g}{4G}{Fourth Generation}
\newacronym{wifi}{Wi-Fi}{Wireless Fidelity}
\newacronym{cran}{C-RAN}{Cloud Radio Access Network}
\newacronym{pusch}{PUSCH}{Physical Uplink Shared Channel}
\newacronym{dmrs}{DM-RS}{Demodulation Reference Signal}
\newacronym{bs}{BS}{Base Station}

\newacronym{snr}{SNR}{Signal to Noise Ratio}
\newacronym{dpoa}{DPoA}{Differential Phase of Arrival}
\newacronym{tdoa}{TDoA}{Time Difference of Arrival}
\newacronym{ota}{OTA}{Over The Air}
\newacronym{los}{LoS}{Line of Sight}
\newacronym{pdf}{PDF}{Probability Density Function}
\newacronym{ppm}{ppm}{Parts per Million}
\newacronym[plural=CIRs,firstplural=Channel Impulse Responses(CIRs)]{cir}{CIR}{Channel Impulse Response}
\newacronym[plural=CFRs,firstplural=Channel Frequency Responses (CFRs)]{cfr}{CFR}{Channel Frequency Response}
\newacronym{nlos}{NLoS}{Non Line of Sight}
\newacronym{mmwave}{mmWave}{millimeter wave}
\newacronym{tx}{TX}{Transmitter}
\newacronym{rx}{RX}{Receiver}
\newacronym{idft}{IDFT}{Inverse Discrete Fourier Transform}
\newacronym{ft}{FT}{Fourier Transform}
\newacronym{ift}{IFT}{Inverse Fourier Transform}
\newacronym{dft}{DFT}{Discrete Fourier Transform}
\newacronym{ofdm}{OFDM}{Orthogonal Frequency Division Multiplexing}
\newacronym{scs}{ScS}{Subcarrier Spacing}
\newacronym{loess}{LOESS}{LOcally Estimated Scatterplot Smoothing}
\newacronym{cdf}{CDF}{Cumulative Distribution Function}
\newacronym[plural=RBs,firstplural=Resource Blocks (RBs)]{rb}{RB}{Resource Block}
\newacronym{mimo}{MIMO}{Multiple Input Multiple Output}
\newacronym{rms}{RMS}{Root Mean Square}
\newacronym{kf}{KF}{Kalman Filter}
\newacronym{iid}{i.i.d.}{Independent and Identically Distributed}
\newacronym{sota}{SOTA}{State-Of-The-Art}
\newacronym{isac}{ISAC}{Integrated Sensing And Communication}
\newacronym{psd}{PSD}{Power Spectral Density}
\newacronym{aoa}{AoA}{Angle of Arrival}
\newacronym{cvm}{CvM}{Cramér--von Mises}
\newacronym{clt}{CLT}{Central Limit Theorem}
\newacronym{wlnn}{WLLN}{Weak Law of Large Numbers}
\newacronym{sar}{SAR}{Synthetic Aperture Radar}
\newacronym{bp}{BP}{Backprojection}

\newacronym{lo}{LO}{Local Oscillator}
\newacronym{if}{IF}{Intermediate Frequency}
\newacronym{rf}{RF}{Radio Frequency}
\newacronym{lna}{LNA}{Low Noise Amplifier}
\newacronym{cfo}{CFO}{Carrier Frequency Offset}
\newacronym{lpf}{LPF}{Low Pass Filter}
\newacronym{hpf}{HPF}{High Pass Filter}
\newacronym{bpf}{BPF}{Band Pass Filter}
\newacronym{p1db}{P1dB}{$1$~dB compression point}
\newacronym{fr4}{FR}{Flame Retardant 4}
\newacronym[plural=PCBs,firstplural=Printed Circuit Boards (PCBs)]{pcb}{PCB}{Printed Circuit Board}
\newacronym{smd}{SMD}{Surface Mount Device}

\newacronym{vna}{VNA}{Vector Network Analyzer}
\newacronym{oslr}{OSLR}{Open, Short, Load, Reciprocal through}

\newacronym[plural=RFSoCs]{rfsoc}{RFSoC}{RF System on a Chip}
\newacronym[plural=NCOs,firstplural=Numerically Controlled Oscillators (NCOs)]{nco}{NCO}{Numerically Controlled Oscillator}
\newacronym{ad}{AD}{Analog to Digital}
\newacronym{da}{DA}{Digital to Analog}

\usepackage{pgfplots}
\usepgfplotslibrary{statistics}
\usepackage{tikz}
\usepackage{pgfplotstable}
\usepackage{xcolor}

\definecolor{mycolor1}{HTML}{003d5c}
\definecolor{mycolor2}{RGB}{103,79,149}
\definecolor{mycolor3}{RGB}{249,89,111}
\definecolor{mycolor4}{RGB}{255,166,0}

\usetikzlibrary{shapes, shapes.misc, arrows, positioning, calc, backgrounds, angles, quotes, patterns, pgfplots.groupplots, arrows.meta, decorations.markings}

\renewcommand{\thetable}{\arabic{table}}

\graphicspath{{./figures/}}
\newcommand{\eq}[1]{Eq.~\eqref{#1}}

\newcommand{\fig}[1]{Fig.~\ref{#1}}
\newcommand{\tab}[1]{Tab.~\ref{#1}}
\newcommand{\secref}[1]{Section~\ref{#1}}

\newcommand{\mytexttilde}{{\raise.17ex\hbox{$\scriptstyle\mathtt{\sim}$}}}

\IEEEaftertitletext{\vspace{-2.\baselineskip}}

\begin{document}
\bstctlcite{IEEEexample:BSTcontrol}
\pagenumbering{gobble}

\title{Beyond Point Targets: Experimental Analysis of Frequency Anisotropy for Multi-band ISAC in FR3}

\author{
\IEEEauthorblockN{
Jacopo Pegoraro\IEEEauthorrefmark{1},
Andrea Bedin\IEEEauthorrefmark{2},
Dario Tagliaferri\IEEEauthorrefmark{3}, Joerg Widmer\IEEEauthorrefmark{2}
}\\
\thanks{Jacopo Pegoraro and Andrea Bedin contributed equally to this work. Corresponding author email: \texttt{jacopo.pegoraro@unipd.it}.

\IEEEauthorrefmark{1}This author is with the University of Padova, Dept. of Information Engineering, Padova, Italy. \IEEEauthorrefmark{2}These authors are with the IMDEA Networks Institute, Madrid, Spain. \IEEEauthorrefmark{3}This author is with the Department of Electronics, Information, and Bioengineering, Politecnico di Milano, Italy.

The project is supported by the Smart Networks and Services Joint Undertaking (SNS JU) and the European Union’s Horizon Europe research and innovation programme under grant agreements 101192521 (Multi-X), 101272485 (PRIME-6G), 101129618 (UNITE) and 101201468 (MiRACLE).}
}

\maketitle

\begin{abstract}
As \gls{isac} systems push toward higher sensing resolution, multi-band processing has emerged as a key enabler, with \gls{fr3} (7-24~GHz) standing out for its combination of wide bandwidth and favorable propagation. 
A common assumption underlying existing multi-band \gls{isac} techniques is that targets behave as frequency-invariant point scatterers, enabling coherent combination of measurements across widely spaced subbands. 
However, this assumption does not hold over wide fractional bandwidths, since real objects exhibit frequency-dependent scattering mechanisms and migrating scattering centers. This paper provides the first systematic experimental characterization of the frequency anisotropy of everyday objects for \gls{isac}, using channel measurements collected with a calibrated vector network analyzer over the 6-24~GHz band, across 10 objects and 120 viewpoints. 
We process bistatic channel impulse responses and synthetic aperture radar images to quantify multi-band coherence, extracting the real part of the cross-band correlation coefficient. 
Our results reveal complex, non-trivial coherence structures that vary substantially with object type and viewing angle, highlighting the need to account for frequency anisotropy in multi-band \gls{isac} system design.
\end{abstract}

\IEEEpeerreviewmaketitle

\section{Introduction}\label{sec:intro}


Multi-band \gls{isac} is attracting attention as a possible solution to the trade-off between spectrum usage and sensing resolution in \gls{6g} systems~\cite{Huawei_multiband,Liu2025_carrieraggregation}.
By coherently combining channel estimates obtained in different and possibly disaggregated frequency bands, existing works have demonstrated sensing resolution gains of up to one order of magnitude, with limited spectrum occupation~\cite{Liu2025_carrieraggregation,Fang2026_carrieraggregation, pegoraro2024hisac}. 

Among the candidate spectra for \gls{6g}, \gls{fr3}, spanning approximately from 7 to 24 GHz, is especially promising for the application of multi-band \gls{isac}~\cite{Manzoni_wavefield, Bomfin2025_multibandFR3channelsensing, Bazzi_compressedsensing_FR3}. 
\gls{fr3} combines favorable propagation characteristics with substantially larger bandwidth than conventional sub-6 GHz systems, while avoiding the severe propagation limitations of millimeter-wave frequencies \cite{Bazzi_uppermidband_COMMAG}.
Moreover, \gls{fr3} potentially offers high \textit{fractional bandwidth}, which is desirable when performing high-resolution sensing based on the \gls{sar} principle~\cite{Manzoni_wavefield}.
Despite these advances, existing multi-band \gls{isac} techniques, including those considering \gls{fr3}, rely on a fundamental assumption that has received little research interest and experimental validation. 
Specifically, targets are typically modeled as collections of \textit{frequency-invariant} point scatterers~\cite{Huawei_multiband, pegoraro2024hisac}, whose complex scattering coefficients remain constant over the considered (wide) bandwidth. 
Under this assumption, measurements acquired at different frequencies can be coherently combined after compensating for propagation delays, enabling synthetic bandwidth expansion and improved range resolution. 
While this approximation is generally adequate for radar or \gls{isac} systems operating over limited fractional bandwidths, its validity becomes questionable when observations span several gigahertz or multiple widely separated frequency bands, as envisioned for \gls{fr3}.
Indeed, real objects are extended electromagnetic scatterers whose dominant scattering mechanisms depend on frequency. 
As the wavelength changes, different physical phenomena--specular reflections, diffraction, penetration, resonance--interact, causing both amplitude and phase of the scattered field to vary across frequency. 
This concept is well known in the \gls{sar} community, where parametric models of the space-frequency back-scattering coefficient of targets have been developed, stemming from the geometrical theory of diffraction \cite{keller2016geometrical,Potter1995GTD}. 
However, the scattering centers' locations are assumed to be \textit{constant} across the whole considered bandwidth. 
This is consistent with the typical \gls{sar} geometry, where the size of the target is much smaller than the propagation distance, and the fractional bandwidth rarely exceeds 10\%. 
Conversely, \gls{isac} systems observe the target from a much closer distance with very different viewpoints, and potentially exhibit very large fractional bandwidth, considering sub-$7$~GHz frequencies, \gls{fr3}, and mmWave bands. 
This causes scattering centers to migrate in space at different frequencies, which generates additional anisotropy.
Although preliminary evidence of frequency anisotropy and the impact of migrating scattering centers has recently been reported in our previous work~\cite{pegoraro2025multibandsensingfr3frequency}, a systematic experimental characterization with calibrated hardware, across different classes of objects, and considering multiple view angles is still lacking.

In this paper, we provide an experimental quantitative assessment of the frequency anisotropy of common objects. 
To this end, we collect an extensive dataset of channel measurements over the $6-24$~GHz frequency range using a calibrated \gls{vna} device.
We consider $10$ different objects, observed from $120$ different viewpoints equally distributed over $360^{\circ}$.
Using such data, we compare bistatic \glspl{cir} and \gls{sar} images obtained using \textit{subbands} with different center frequencies. 
Both the \glspl{cir} and the images are processed by removing the propagation carrier phase using the \gls{bp} algorithm~\cite{Manzoni_wavefield}, thus isolating the complex-valued reflectivity of the targets across frequencies.

Our analysis evaluates the \textit{coherence} of basic shapes, such as spheres and cubes, and everyday objects, such as tables and chairs.
In this paper, coherence is a measure of the \textit{magnitude and phase} discrepancy between the target's contribution to \glspl{cir} or images in different subbands, and it is computed as the \textit{real part} of the common coherence metric used in \gls{sar}~\cite{touzi1999coherence}.
We provide a thorough statistical analysis of coherence across different viewpoints of the targets, and showcase the significant variation of coherence depending on the specific pair of considered subbands.

The main contributions of this paper are summarized next.

\begin{itemize}[leftmargin=*]
    \itemindent=1mm
    \item We present the first extensive experimental analysis of frequency anisotropy for multi-band \gls{isac} in \gls{fr3}. We collect channel measurements using calibrated hardware over the $6-24$~GHz frequency band for $120$ different viewpoints on the objects, providing an exhaustive characterization of their scattering properties across frequency.
    The dataset will be openly released to the community.
    \item We propose a methodology to evaluate the coherence of the scattering coefficients of the objects. After carrier phase compensation, we propose to compute the real part of the standard coherence metric used in the literature. 
    This allows capturing both amplitude and phase correlations across subbands, and computing cross-subband coherence matrices to highlight coherence patterns of specific objects.
    \item We derive new insights on frequency anisotropy of everyday objects compared to simple shapes. The multi-band coherence shows highly non-trivial correlation patterns across frequencies, and can change from perfect coherence to complete incoherence for different view angles.
    This underscores the key importance of taking frequency anisotropy into account when designing multi-band \gls{isac} systems.   
\end{itemize}


\section{Example of frequency anisotropy}
\label{sec:example}

\begin{figure}[t!]
\centering
\includegraphics[width=\linewidth]{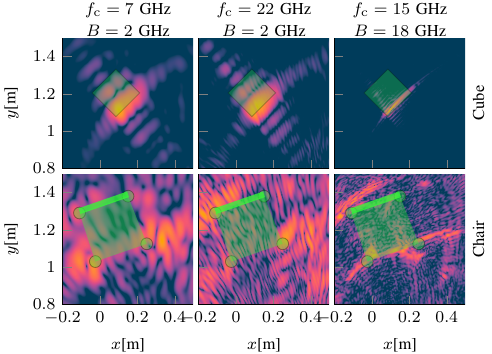}
\vspace{-3mm}
\caption{Example images. The object's silhouette is represented by the green areas. This is approximate and for illustration purposes only.}
\label{fig:img_examples}
\end{figure}

To demonstrate frequency anisotropy, we provide an example from the dataset used in this study. 
\fig{fig:img_examples} shows \gls{sar} images obtained using the \gls{bp} algorithm on a $1.1$~m synthetic array aperture with $2$~cm antenna spacing with $B=2$~GHz bandwidth centered around $f_{\rm c}=7$~GHz~(left), with $B=2$~GHz bandwidth centered around $f_{\rm c}=22$~GHz~(center), and with the full $B=18$~GHz available bandwidth~(right). 

The top row shows the images of a simple cube with a metallic surface. 
The silhouette of the cube in two dimensions is highlighted in green. 
In the full band image, the face of the cube is clearly visible, while the rest of the image shows low amplitude. 
As expected, images made with $2$~GHz of bandwidth present a much wider main peak, as well as sidelobes due to the limited bandwidth. 
The images at $f_{\rm c}=7$~GHz and $f_{\rm c}=22$~GHz are similar, suggesting that the cube is a fairly isotropic object when viewed from this specific frontal angle.
As we will show in \secref{sec:image-results}, this is not the case for all possible rotation angles of the cube. 

Conversely, the second row of images represents a chair, with a top-down view superimposed in green. 
In this case, the images realized with different carrier frequencies show significant differences, and the full-band image does not show the same sharp contrast as with the cube. 
This suggests that the chair exhibits significant frequency anisotropy, due to changes in the amplitude and phase of the scattering coefficients of the object across frequency.

The above example motivates further analysis of the multi-band coherence of everyday objects with quantitative methods, as done in the following sections.

 \section{Experimental setup}
\label{sec:expsetup}

\begin{figure*}[t!]
\centering
\includegraphics[width=\linewidth]{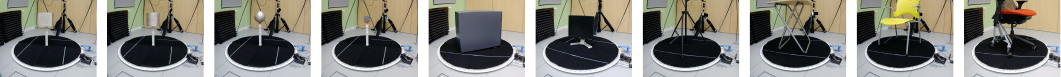}
\caption{Objects used in the experiments. From left to right: cube, cylinder, sphere, plate, cabinet, monitor, tripod, table, chair, office chair.}
\label{fig:obj_pics}
\vspace{-2mm}
\end{figure*}

\begin{table}[t!]
\small
\centering
  \setlength\extrarowheight{-6pt}
  \caption{Object's materials and dimensions.} \label{tab:objects}
  \vspace{-2mm}
\begin{tabular}{c c c}
\toprule
\textbf{Object} & \textbf{Material} & \textbf{Size~[cm]}  \\
\midrule
basic shapes & metalized 3D print & $10$-$20$ \\
cabinet, tripod & metal & $70$ \\
monitor & various materials & $45$\\
chair, office chair & metal, plastic & $60$ \\
table & metal, wood & $100$\\
\bottomrule
\end{tabular}
\vspace{-2mm}
\end{table}

\begin{figure}[t]
\centering
\includegraphics[width=0.5\linewidth]{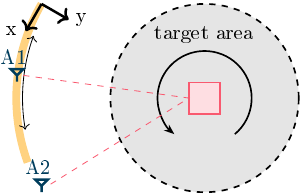}
\caption{Experimental setup.}
\label{fig:setup}
\vspace{-3mm}
\end{figure}

Measuring magnitude and phase variations in the scattering coefficient of a sensing target across a wide frequency range is experimentally challenging and requires calibrated hardware to avoid artifacts related to hardware non-idealities.

\textbf{Measurement device and calibration.} We utilize a N9952B Fieldfox \gls{vna} connected to two Vivaldi antennas, measuring the $\text{S}_{21}$ parameter, which corresponds to the bistatic \gls{cfr} between the two antennas, across the $6$~GHz to $24$~GHz band. 
We calibrate the \gls{vna} at the antenna connector plane with an \gls{oslr} calibration~\cite{cal_info} using a SCKCL26-3.5 mechanical calibration kit~\cite{cal_kit}, compensating for any effect introduced by the cables and instrument non-idealities. 
Secondly, we measure the bistatic \gls{cfr} with the antennas facing each other at a known distance ($850$~mm), with anechoic panels placed around the antennas to mitigate multipath. 
In this setting, as we calibrated the \gls{vna}, we measure the free-space \gls{los} path and the antenna response. 
By compensating for the free-space \gls{los} path, we obtain an estimate of the antenna response that is subsequently compensated for in all measurements, as described in~\secref{sec:method}.

\textbf{Targets and experimental setup.} We consider $10$ different target objects, shown in~\fig{fig:obj_pics}. The materials and approximate sizes (largest dimension in the horizonal plane) of the objects are listed in \tab{tab:objects}.
These include $4$ simple shapes (cube, cylinder, sphere, and flat plate, from left to right) that are 3D-printed and coated with a conductive paint~\cite{conductive_paint}, and $6$ real-life objects (cabinet, monitor, tripod, table, a chair, and an office chair, from left to right). 

The experimental setup is sketched in \fig{fig:setup}. The \gls{vna} has two channels, denoted by channel~$1$ and channel~$2$.
Channel~$2$ is connected to a fixed antenna A2, and channel~$1$ is connected to antenna A1, which can be moved on a circular rail of radius $1.2$~m (depicted in yellow). The spherical rail has been chosen to ensure that the antenna boresight is always facing the object, to minimize the impact of the antenna radiation pattern.
Exploiting the different locations of antenna A1 we create a synthetic spatial aperture that is used to generate the~\gls{sar} images.
The target object is placed in the center of a rotating disk (the grey area in~\fig{fig:setup}) used to obtain channel measurements at different viewpoints on the target.
The surroundings of the measurement area are covered with panels made of radio-frequency-absorbing material to attenuate the contribution of multipath components due to objects and walls in the background.

\textbf{Measurement acquisition protocol.} During each acquisition, we move antenna A1 across $N_{\rm ant} = 56$ positions on the rail, spaced by $2$~cm and spanning an aperture of $1.1$~m. 
For each position of antenna A1 on the rail, we rotate the objects in steps of $3^\circ$ covering $N_{\rm ang} = 120$ angles across the $360^{\circ}$ possible viewpoints.
In each combination of antenna position on the rail and object rotation, we measure the \gls{cfr} across $K=801$ frequency samples spanning the $6-24$~GHz band, for a total of $B_{\rm tot}=18$~GHz of bandwidth, where $F=22.5$~MHz is the frequency sampling period.
Under these conditions, the carrier frequency of the acquisitions is $f_0 = 15$~GHz.
We denote the \gls{cfr} measured with an object oriented at the $m$-th angle, with antenna A1 in the $i$-th position at the $k$-th frequency, by $H_{i,m}(f_0 + kF)$.
Given the considered experimental setup, the antenna position index is $i=1, \dots, N_{\rm ant}$, the object rotation index is $m = 1,...,N_{\rm ang}$, and the frequency index is $k = -\lfloor K/2 \rfloor, \dots, \lfloor K/2 \rfloor$.

 \section{Multi-band coherence evaluation}
 \label{sec:method}

To evaluate the frequency anisotropy of objects, we analyze the correlation among \glspl{cir} and among \gls{sar} images generated using subbands of the full \gls{cfr}.
In the following, we present the preprocessing steps needed to obtain the \gls{cir} for each subband, the \gls{bp} algorithm to produce the \gls{sar} images, and we detail our method to evaluate the coherence of the targets. 

We describe our methodology using continuous frequency and delay variables, although the experimental data discretizes frequency as described in~\secref{sec:expsetup}.
This simplifies the presentation without changing the underlying concepts.
The full band \gls{cfr} obtained with antenna position $i$ and object rotation $m$ is thus $H_{i,m}(f)$, with $f\in [f_0 - B_{\rm tot}/2, f_0 + B_{\rm tot}/2]$.

\subsection{Preprocessing steps}\label{sec:preprocessing}

{\bf Antenna response removal.} The full band \gls{cfr} contains the contribution of the Vivaldi antennas used in the measurements.
This response is not ideal, so it may affect the evaluation of the target coherence if not properly compensated for.
As detailed in \secref{sec:expsetup}, the frequency response of the Vivaldi antennas is estimated to allow removing its contribution from the \gls{cfr}.
Denote the estimated antenna response by $H_{\rm ant}(f)$.
This is used to obtain a compensated \gls{cfr} as $\bar{H}_{i,m}(f) = H_{i,m}(f) / H_{\rm ant}(f)$,
which contains the sole channel response.

{\bf Target contribution extraction.} $\bar{H}_{i,m}(f)$ contains the contribution of the target of interest and that of the other multipath components caused by the direct propagation from antenna A1 to A2 and the residual contribution of surrounding objects.
To isolate the target's contribution, the setup is constructed to ensure that the target lies within a bistatic range interval $\mathcal{R} = [1.5, 3.5]$~m, whereas any other multipath reflector is over $4$~m away. 
Thus, we exploit the very wide bandwidth ($18$~GHz) of the full \gls{cfr} $\bar{H}_{i,m}(f)$, which has a centimeter-level resolution, to isolate the target's contribution.
More specifically, defining the full-band \gls{cir} as $h_{i,m}(\tau) = \mathrm{IFT}\left\{\bar{H}_{i,m}(f) \right\}$, in the following we will consider the time-gated \gls{cfr} $\widetilde{H}_{i,m}(f) = \mathrm{FT}\{\mathbf{1}_{\mathcal{R}}(\tau) h_{i,m}(\tau)\}$, where $\mathbf{1}_{\mathcal{R}}$ is the indicator function of the delay interval corresponding to $\mathcal{R}$.

{\bf Subband-specific \gls{cir} computation.} We consider subbands of equal bandwidth $B$, where the $n$-th considered subband has center frequency $f_{n}$. 
Denote the \gls{cfr} of the $n$-th subband with bandwidth $B$, with the $i$-th antenna position and $m$-th object rotation by 
\begin{equation}
    H^{(n)}_{i,m}(\nu) = \widetilde{H}_{i,m}(f_n + \nu), \quad \nu\in \left[- \frac{B}{2F},  \frac{B}{2F} \right] \mbox{Hz},
\end{equation}
where $\nu$ is a relative frequency to the center frequency of the subband.
The bandwidth-limited \gls{cir} is obtained via \gls{ift} from $H^{(n)}_{i, m}(\nu)$ as
\begin{equation}
    h^{(n)}_{i,m}(\tau) = \mathrm{IFT}\left\{H_{i,m}^{(n)}(\nu) \right\},
\end{equation}
where $\tau$ is the propagation delay variable.

To exemplify how the \glspl{cir} of the subbands are used to evaluate the target's coherence, let us consider a frequency-isotropic point-like target with delay $\tau_{0}$, with a scattering coefficient having magnitude $\alpha$ and phase $\psi$.
This target would lead to a \gls{cir} equal to
\begin{equation}
    h^{(n)}_{i,m}(\tau) = \alpha e^{j \psi}B\mathrm{sinc}(B (\tau - \tau_0)) e^{-j 2 \pi f_n \tau_0},\label{eq:examplecir}
\end{equation}
where the carrier phase term due to the propagation delay, $2 \pi f_n \tau_0$, prevents a direct evaluation of the scattering coefficient $\alpha e^{j \psi}$ across frequency.
Hence, to eliminate this phase term, we operate similarly to the \gls{bp} algorithm, which computes the phase-compensated \gls{cir}, defined as 
\begin{equation}
    \widehat{h}_{i,m}^{(n)}(\tau) =  h^{(n)}_{i,m}(\tau) e^{j 2 \pi f_n\tau}\label{eq:phasecompcir} 
\end{equation}
The above operation compensates for the carrier phase at $\tau=\tau_0$, so that in the example shown in \eq{eq:examplecir}, $\widehat{h}_{i,m}^{(n)}(\tau_0) =  \alpha e^{j \psi} B$, which is a rescaled version of the target's scattering coefficient.
Thus, a point-like frequency-isotropic target implies the consistency of $\widehat{h}_{i,m}^{(n)}(\tau_0)$ over frequency.
This allows evaluating the coherence of a target by testing whether the phase of the compensated \gls{cir} is constant across frequency. 
In the following sections, we use the phase-compensated \gls{cir} to obtain images of the target and its coherence. 

{\bf Formation of images via \gls{bp}.} The \gls{bp} algorithm synthesizes a spatial aperture by coherently combining the phase-compensated \glspl{cir} obtained at different locations of antenna A1.
The combination is performed by directly summing the \glspl{cir} from all the different antenna locations, $i=1,\dots, N_{\rm ant}$.
To obtain a 2D image, the bistatic delay variable in the \gls{cir} is parametrized by an image pixel, whose location on the horizontal $x, y$ plane is represented by vector $\mathbf{x}\in \mathbb{R}^2$.
Similarly, the known locations of antennas A1 and A2 are denoted by $\mathbf{x}^{\rm A1}_i$ and $\mathbf{x}^{\rm A2}$, respectively, with $\mathbf{x}^{\rm A1}_i$ being dependent on the location index $i$.
We denote the parametrized bistatic delay by $\tau_i(\mathbf{x}) = (\|\mathbf{x} -\mathbf{x}^{\rm A1}_i\|+\|\mathbf{x} -\mathbf{x}^{\rm A2}\|)/c$ where $c$ is the speed of light.
The image of the target using subband $n$ at viewpoint $m$ is obtained as~\cite{Manzoni_wavefield}
\vspace{-2mm}
\begin{equation}\label{eq:bp-image}
    I^{(n)}_{m}(\mathbf{x}) = \sum_{i=1}^{N_{\rm ant}} \widehat{h}_{i,m}^{(n)}(\tau_i(\mathbf{x})),
\end{equation}
where each pixel's value represents the complex-valued reflectivity of the corresponding point in space.

\subsection{Coherence metrics}
{\bf \gls{cir} coherence metric.} To evaluate the discrepancy between \glspl{cir} obtained in different subbands, we take the real part of the complex-valued coherence metric used, e.g., in~\cite{touzi1999coherence}.
More specifically, the real part of the coherence between the \glspl{cir} of subbands $n_1$ and $n_2$ at antenna location $i$ and object orientation $m$ is defined as
\begin{equation}\label{eq:coherence-cir}
    \gamma^{\rm CIR}_{i,m}(n_1, n_2) = \frac{ \Re \left\{\mathbb E_{\mathcal{T}_{i,m}}\left[  \widehat{h}^{(n_1)}_{i,m}(\tau)\left(\widehat{h}^{(n_2)}_{i,m}(\tau)\right)^*\right]\right\}}{\sqrt{\mathbb E_{\mathcal{T}_{i,m}}\left[\left|\widehat{h}_{i,m}(\tau)\right|^2\right] \mathbb E_{\mathcal{T}_{i,m}}\left[\left|\widehat{h}_{i,m}(\tau)\right|^2\right]}},
\end{equation}
where $\Re\{a\}$ is the real part of $a$ and $\mathbb E_{\mathcal{T}_{i,m}}$ is the expectation over the delay variable $\tau\in\mathcal{T}_{i,m}$, and 
\begin{equation}
    \mathcal{T}_{i,m} = \left\{\tau: \left|\widetilde{h}_{i,m}(\tau) \right| > \eta \max_{\tau}\left|\widetilde{h}_{i,m}(\tau) \right|\right\}
\end{equation}
is the set of delays where the full-band \gls{cir} $\widetilde{h}_{i,m}(\tau)$ is above $\eta$ times its maximum, taken as an approximation for the set of delays where a target is present (the "$\tau_0$"s of \eq{eq:examplecir}).
In our results, we use \mbox{$\eta=0.7$}, based on an empirical evaluation of the coherence values.
Further, note that to coherently combine two \glspl{cir} at different subbands, their \textit{phases need to align} for each delay value. 
This only happens when the cross-correlation has a large, \emph{real and positive} value, making the use of the real part, instead of the more commonly used magnitude, critical.

{\bf Image coherence metric.} The image coherence is defined similarly to the \gls{cir} coherence as
\begin{equation}
    \gamma^{\rm I}_{m}(n_1, n_2) = \frac{\Re \left\{\mathbb E_{\mathcal{X}_{m}}\left[I^{(n_1)}_{m}(\mathbf{x})\left(I^{(n_2)}_{m}(\mathbf{x})\right)^*\right]\right\}}{\sqrt{\mathbb E_{\mathcal{X}_{m}}\left[\left|I^{(n_1)}_{m}(\mathbf{x})\right|^2\right] \mathbb E_{\mathcal{X}_{m}}\left[\left|I^{(n_2)}_{m}(\mathbf{x})\right|^2\right]}},
\end{equation}
where $\mathcal{X}_{m}$ is a set of pixels selected using a similar reasoning to the \gls{cir} case, i.e., 
\begin{equation}
   \mathcal{X}_{m} = \left\{\mathbf{x}: \left|I_{m}(\mathbf{x}) \right| > \eta \max_{\mathbf{x}}\left|I_{m}(\mathbf{x}) \right|\right\},   
\end{equation}
with $I_{m}(\mathbf{x})$ being the full band image obtained via \gls{bp} using the full band \glspl{cir}, $\widetilde{h}_{i,m}(\tau)$, in \eq{eq:bp-image}.
The coherence between images does not depend on the antenna position $i$ since all antenna positions are used in the image computation (\eq{eq:bp-image}).

In \secref{sec:results}, we report the values of $\gamma^{\rm CIR}_{i,m}$ or $\gamma^{\rm I}_{m}$ depending on whether \glspl{cir} or images are considered.

\section{Results}\label{sec:results}

In this section, we present the obtained coherence results on the \glspl{cir} and images using the previously described methodology.
We divide the discussion into \gls{cir} anisotropy evaluation and image anisotropy evaluation.
For each case, we present the coherence distribution for each considered object over the $N_{\rm ang}=120$ rotation angles.

In our results, we select a total of $N=31$ partially overlapping subbands of bandwidth $B=2$~GHz, with center frequencies \mbox{$f_n = 7 + 0.5(n-1)$~GHz}, $n=1, \dots, N$. 
We leave a comprehensive analysis of the impact of the subbands' bandwidth $B$ for future work.

\subsection{\gls{cir} anisotropy}
In this section, we present the \gls{cir} anisotropy results.
Analyzing coherence based on \glspl{cir} gives insight into how anisotropy affects multi-band combination when channel estimates are only obtained from a single location of the bistatic \gls{isac} devices.
In this case, obtaining an image of the target is not feasible, and multi-band methods are typically applied to compute a high-resolution \gls{cir} estimate~\cite{pegoraro2024hisac}.
However, coherently combining subbands across which the target is incoherent does not increase the sensing resolution but rather degrades the resulting \gls{cir} estimate~\cite{pegoraro2025multibandsensingfr3frequency}.
When considering \gls{cir} coherence, we select a single location of antenna A1, i.e., the one closest to A2 in~\fig{fig:setup}, denoting it by index $i=1$.
This choice is made to isolate the impact of the object rotation, leaving the study of antenna location impact for future work.

\begin{figure}[t!]
\centering
\includegraphics[width=\linewidth]{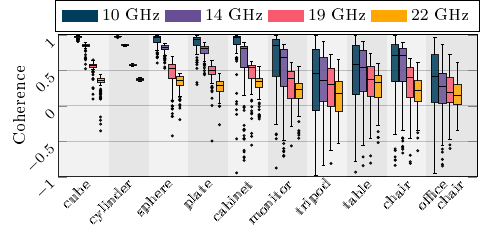}
 \vspace{-2mm}
\caption{\gls{cir} coherence distribution for different view angles.}
\label{fig:boxplot_cirs}
\end{figure}

{\bf Coherence distribution across different views.} 
 In \fig{fig:boxplot_cirs}, we show a boxplot of the distribution of the coherence $\gamma^{\rm CIR}_{1,m}(n_1, n_2)$ over $m=1, \dots, N_{\rm ang}$ and all the $10$ objects from \fig{fig:obj_pics}, for $f_{n_1}=7$~GHz and $f_{n_2} \in \{10, 14, 19, 22\}$~GHz.

The $4$ simple objects, namely cube, cylinder, sphere, and plate, all exhibit fairly high coherence, with a median over $0.5$ for all center frequencies $f_{n_2}$.
Nevertheless, a coherence degradation for higher center frequencies is observed, and the cylinder shows a significantly lower coherence variance across view angles compared to the other simple objects.
While this is expected from the cube and plate, which are theoretically anisotropic with respect to the view angles, it is interesting to notice when considering the sphere.
This is likely due to the lower reflectivity of the sphere compared to the cylinder due to its smaller dimension, which makes its coherence evaluation more affected by noise.

The everyday objects, namely from cabinet to office chair, all exhibit significantly different coherence values depending on the view angle, which results in a large variance in the boxplot distribution.
This demonstrates that complex objects have very different scattering behavior when observed from different view angles, even at similar frequencies, so multi-band coherent combination in \gls{isac} should be applied with care.
Moreover, some objects like the chair and the office chair show a \textit{non-monotonically decreasing} coherence as a function of frequency.
For example, the median coherence of the chair with respect to the reference $f_{n_1}=7$~GHz subband is higher at $f_{n_2}=14$~GHz than at $f_{n_2}=10$~GHz.
This is a non-trivial behavior which suggests complex scattering phenomena occur with everyday objects.

\begin{figure}[t!]
\centering
\includegraphics[width=\linewidth]{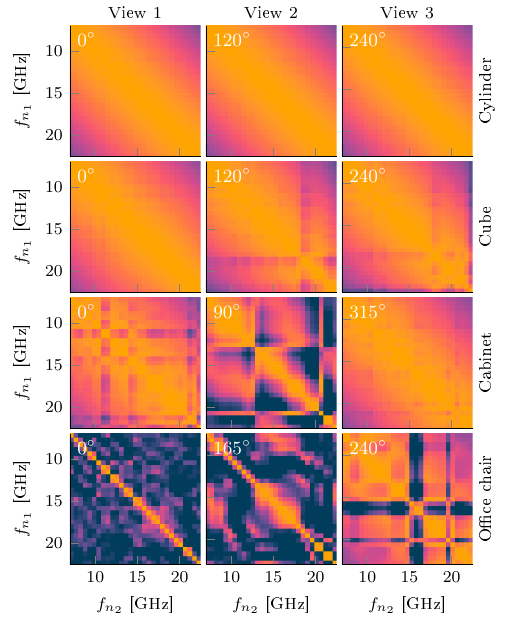}
 \vspace{-2mm}
\caption{Examples of \gls{cir} coherence for three representative views of different targets. Lighter colors correspond to high coherence, darker colors to low coherence.}
\label{fig:camt_objs}
\end{figure}

{\bf Cross-band coherence analysis.} To get a deeper insight into the variations of coherence as a function of center frequency, in \fig{fig:camt_objs} we report the cross-band coherence values for the cylinder, cube, cabinet, and office chair, for all combinations of $n_1, n_2 \in \{1, \dots, N\}$.
The $3$ columns refer to different representative view angles~$m$, and lighter/darker colors refer to high/low coherence, respectively.

We observe that the cylinder and the cube have high coherence for most of the subband combinations, which decreases for very distant bands.
While this behavior is consistent across views for the cylinder, some slight variations are present for the cube.
Conversely, the cabinet and the office chair show
complex coherence structure across different subband pairs, for some views, with non-monotonic, periodic, or cluster-like patterns.
Such structure also changes completely for different view angles on the same object, demonstrating the importance of frequency anisotropy in multi-band coherent combination.

\subsection{Image anisotropy}\label{sec:image-results}

In this section, we present the image anisotropy results.
This evaluation is required to highlight frequency anisotropy effects in relation to coherent combination across space, i.e., different locations of antenna A1.
Indeed, different from airborne \gls{sar} systems, the antenna locations used in \gls{isac} are usually not under control, since they belong to static transceivers such as base stations/access points or users.
Antenna locations are the same for all the considered frequency bands, leading to different spatial sampling artifacts, e.g., grating lobes, at different frequencies.
Moreover, images are also more affected by the migration of the scattering centers in space compared to \glspl{cir}, since the object is observed from multiple locations.

\begin{figure}[t!]
\centering
\includegraphics[width=\linewidth]{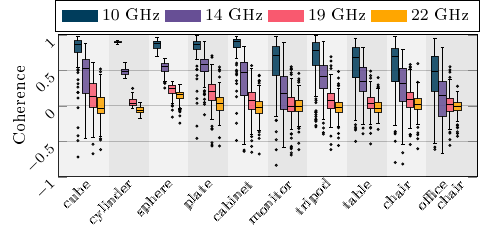}
 \vspace{-2mm}
\caption{Image coherence distribution for different view angles.}
\label{fig:boxplot_images}
\end{figure}

 {\bf Coherence distribution across different views.} In \fig{fig:boxplot_images}, we show the distribution of the image coherence, $\gamma^{\rm I}_{m}$ across different object rotation angles~$m$. 
The first difference with respect to the \gls{cir} coherence in \fig{fig:boxplot_cirs} is the overall lower coherence of the images, which decays to a median around $0$ at $22$~GHz for all objects. 
This can be explained by noticing that for the very wide range of considered frequencies, the antenna positions spacing of $2$~cm corresponds to super-wavelength spatial sampling for center frequencies higher than $15$~GHz, which introduces grating lobes in the images.
Such grating lobes can degrade the coherence, but they represent an unavoidable phenomenon in multi-band \gls{isac} systems where spatial sampling locations are fixed and not under control.

Secondly, the cylinder and the sphere exhibit much lower coherence variance compared to other objects, due to their radial symmetry, while everyday objects show lower and more variable coherence as frequency increases.

\begin{figure}[t!]
\centering
\includegraphics[width=\linewidth]{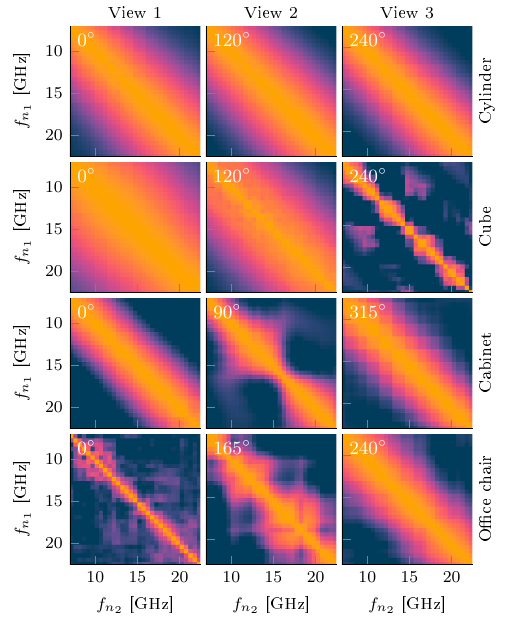}
 \vspace{-2mm}
\caption{Example of image coherence for complex objects. Lighter colors correspond to high coherence, darker colors to low coherence.}
\label{fig:camt_imgs}
\end{figure}

{\bf Cross-band coherence analysis.} In \fig{fig:camt_imgs}, we report the cross-band coherence for the cylinder, cube, cabinet, and office chair, considering all possible combinations of $n_1, n_2 \in \{1, \dots, N\}$.
The $3$ columns refer to different view angles~$m$.

The cylinder and cube mostly show high coherence, although this decays for distant frequency bands.
Notably, at specific view angles such as view~$3$, the cube becomes almost completely incoherent, due to its highly directional scattering response that emerges when considering spatial images.
 
Conversely, the cabinet and office chair show a significantly different behavior.
First, the image coherence is always low for distant subbands, and there are no periodic patterns.
However, for some views we observe clusters of closely spaced subbands that preserve high coherence across up to $4-5$~GHz of bandwidth (view $2$), possibly offering room for significant resolution improvement via multi-band combination.
For other view angles instead, e.g., view~$1$ for the office chair, the object almost always exhibits low coherence.

\section{Concluding remarks}\label{sec:conclusion}

In this paper, we provide a systematic experimental analysis of frequency anisotropy of common objects for \gls{isac} in \gls{fr3}. 
We note how larger, more complex objects exhibit a more notable and more complex anisotropic behavior than small, simple shapes. 
Further, we see how the anisotropy can significantly differ depending on the angle at which the object is observed, and on whether it is evaluated on a single \gls{cir} or a \gls{sar} image. 
We also observe that the decorrelation is not always monotonic with the difference between the considered frequencies, and it can exhibit complex patterns where the correlated frequencies appear periodically or in clusters.  

\bibliography{references.bib}
\bibliographystyle{IEEEtran}
\end{document}